# Visualization of Clandestine Labs from Seizure Reports: Thematic Mapping and Data Mining Research Directions

William H. Hsu[1]
bhsu@ksu.edu

Mohammed Abduljabbar[1]
xec@ksu.edu

Ryuichi Osuga[1]
ryusuga@ksu.edu

Max Lu[2]
maxlu@ksu.edu

Wesam Elshamy[1]
welshamy@ksu.edu

[1] Department of Computing and Information Sciences
[2] Department of Geography
Kansas State University
Manhattan, KS 66506
+1 785 236 8247

## ABSTRACT
The problem of spatiotemporal event visualization based on reports entails subtasks ranging from named entity recognition to relationship extraction and mapping of events. We present an approach to event extraction that is driven by data mining and visualization goals, particularly thematic mapping and trend analysis. This paper focuses on bridging the information extraction and visualization tasks and investigates topic modeling approaches. We develop a static, finite topic model and examine the potential benefits and feasibility of extending this to dynamic topic modeling with a large number of topics and continuous tome. We describe an experimental test bed for event mapping that uses this end-to-end information retrieval system, and report preliminary results on a geoinformatics problem: tracking of methamphetamine lab seizure events across time and space.

## Categories and Subject Descriptors
H.3.3 [**Information Storage and Retrieval**]: Information search and retrieval – *clustering, relevance feedback, selection process*; H.2.8 [**Database Management**]: Database applications – *data mining, spatial databases and GIS*

## 1. INTRODUCTION
In this paper, we address the problem of event visualization based on structured data, in the form of time-referenced and georeferenced relational tuples, and on unstructured data, in the form of free text. Information extraction systems based on named entity recognition (NER) and relationship extraction have enabled detection of events mentioned in free text and extraction of structured tuples describing the location, time, along with other attributes of an event. Identifying hotspots and trends, however, remains an open problem. One limitation is the absence of ground truth for high event activity. In some cases this is due to a lack of well-defined criteria for activity and relevance, while in some it is due to limitations in existing annotation interfaces.

We first present a basic approach to event visualization. Our general framework makes use of mapping tools such as *Google Maps* [1], the Google web toolkit, and timeline visualization tools such as *MIT SIMILE* [2]. It also builds upon previous work on gazetteer-based event recognition and syntactic patterns for semantic relationship detection. Next, we show how a system developed originally for visualization of animal disease outbreaks reported in online news documents can be adapted to display reports of methamphetamine lab seizures compiled by regional law enforcement. We briefly outline the development of a domain-specific data description language for increased portability and ease of information integration. We then discuss the role of topic modeling and information retrieval approaches in filtering and ranking events.

A key technical contribution of this work is the application of topic modeling algorithms in order to compute the posterior probability of a particular spatial location, time unit, or combination given the type of event, which is treated as a topic. This allows the data to be interrogated systematically in order to display geographic regions that are more prone to events of interest. A potential application of this is to construct a time composite map of administrative divisions within a state or province, or a spatial composite time series by month or year, showing active regions. These can be visualized using a *choropleth map*: a map in which regions (geographic regions in this case) are coded by colors or grayscale intensity levels. These represent a variable of interest – in this case, event frequency. Finally, the ability to estimate marginal likelihoods over locations and times given the event type parameters can also be used to filter events, to display only those that fall within a specified frequency range. For example, the system can be configured to search for seizures of methaphetamine production labs in counties or districts where they are common or rare.

## 2. EVENT VISUALIZATION TASKS
### 2.1 Spatiotemporal Event Extraction
The goal of event extraction is to identify phenomena related to specific actions, occurrences, relationships, or entities. For example, a positive test for a contagious animal disease on a farm is an event that may be tied to an epidemic and identified *post hoc* as indicating an outbreak. The seizure of equipment from a methamphetamine production facility, or of waste products from a dump site, is an example of an event in the domain of drug enforcement. Events that can be localized in space and time form the basis of spatiotemporal event extraction.

In the domains of veterinary epidemiology and drug enforcement, decision support systems are typically based on spatiotemporal event extraction and visualization. When events are already available in structured form, they are usually compiled manually from investigative reports by local or national authorities: animal health agencies in the case of veterinary epidemiology and state of national bureaus of investigation in the case of drug seizures. By contrast, unstructured data often comes into the decision support system as the result of a web crawl based on domain-specific resource identifiers: seeds (URLs) and search terms. Federated displays and user interfaces for these decision support systems often combine event data from structured data repositories with data extracted from free text. This entails data integration challenges such as disambiguation, deduplication and identity uncertainty for entities and events; expansion of existing named entity sets from gazetteers (lists of known entities); and inference of attributes for relationships representing events and entities representing actors and objects. These topics are beyond the scope of this paper; we refer the interested reader to existing literature on the current state of the field in domain-focused relationship extraction.

Instead, we suppose here that some preliminary classification step has already taken place to identify the entity that serves as anchor point for an event, and that further classification or inference has identified a putative time and location for the event. Whether this is accomplished through supervised inductive learning from text corpora or as a result of basic pattern matching, our starting point is a candidate tuple to be analyzed, considered for presentation to a decision-maker or search user, and if selected, visualized in the context of events of interest.

## 2.2 Georeferencing and Map Visualization

Mapping out spatially-referenced events, even using structured data sources, entails a straighforward but data-intensive georeferencing task: looking up the coordinates (latitude and longitude) of street addresses and postal codes where events are reported to have occurred.

The resulting coordinates are placed into a spatial database management system (SDBMS) for visualization using software libraries and services such as *Google* Maps, as shown in Figure 1. For this purpose, we developed two alternative access layers with a unified representation and geographic information system (GIS) data model. The first layer is based on Google's Keyhole Markup Language (KML) and a file-based application programmer interface (API), while the second layer is based on a PHP interface to a MySQL database implementing the KML schema. Our front-end application, *TimeMap*, can be configured to use either layer.

## 2.3 Timeline Visualization

Figure 1 depicts the data integration between the map and timeline visualization subsystems. The seizure event in April, 2010 is represented on the map by a pop-up note, on the monthly scale timeline (upper right) by a circled dot, and on the yearly scale timeline (lower right) by a circled point.

## 2.4 Thematic Mapping

The object of thematic mapping is to depict phenomena and trends in a geospatial context. Toward this end, we have added a mapping overlay to the *TimeMap* framework that allows map transformation such as superimposition of data and transparency layers to be applied using the Google Maps API. This includes choropleth maps with dynamically computable color palettes.

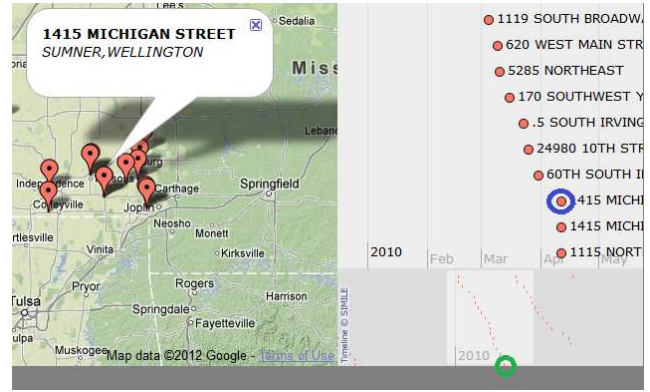

**Figure 1. Map and timeline visualization of meth lab seizure events (2004-2011) using *Google Maps* and *MIT SIMILE*. Seizures from the first half of 2010 are depicted, with one event selected.**

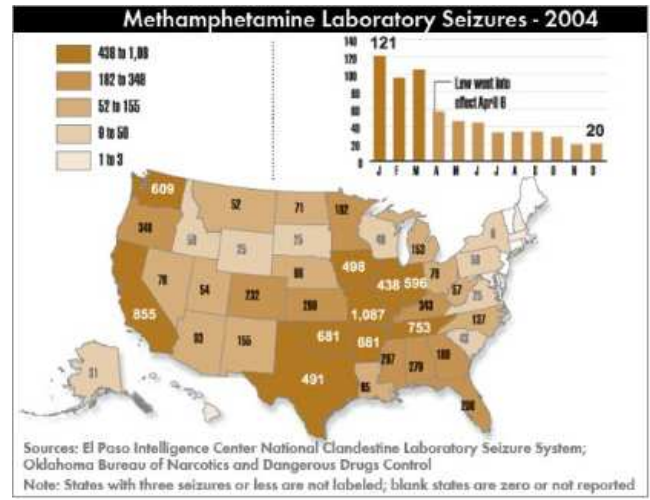

**Figure 2. Choropleth map of 2004 meth lab seizures in the USA. (Associated Press, 2005)**

Figure 2 is a choropleth map depicting meth lab seizures by state in 2004. [3] This map, which was published by the Associated Press and featured on the ABC News web site in 2005, does not take state population or intrastate population distribution into account. More importantly for information retrieval applications, it does not provide any drill-down interface *cf. HealthMap* [4] or similar event visualization services. One of the reasons for the development of the geospatial visualization components of *TimeMap* was to facilitate information retrieval and multimodal information access using well-established visualization techniques such as thematic mapping and small multiples.

## 2.5 Spatial Time Series Prediction

A final rationale for the *TimeMap* visualization framework arises from domain-specific data mining objectives in epidemiology and criminology. Governmental agencies devoted to agriculture, public health, and law enforcement often encounter a need for predictive analytics tools to assist with decision-making in both public policy and intervention, and with civic outreach. In the domain of public health, tools such as *HealthMap* [4] have begun to do for individual citizens what more general crime-mapping systems are intended do for search users: provide relevance filters based on criteria related to incident frequency, corroborative reporting, and significance.

## 3. TOPIC MODELING

As mentioned in Section 2.1, named entity recognition combined with date and location can provide a means of extracting a stream of events and updates from news stories. This also holds for microblogs and other social media. In order to classify new events and detect the emergence or **revival** of new event-related topics, however, a mechanism for monitoring update streams is needed. This requires a more flexible topic model than the fixed or expandable sets of named entities used for structured information extraction. Furthermore, the frequency and semantic heterogeneity of event reporting from multiple media outlets, even on the web alone, may require enrichment of the parameters beyond those used in classical generative models for information retrieval. We now examine possible extensions to these document clustering models.

## 3.1 Static Topic Models

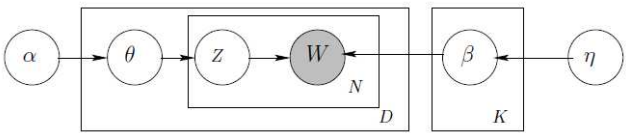

**Figure 3. Plate model for Latent Dirichlet Allocation (LDA) in a system with an *N*-word lexicon, *D* documents, and *K* topics.**

Figure 3 illustrates the kind of generative Bayesian topic model widely used to cluster static collections of documents. Here, $\theta$ is a topic distribution for a document, while $Z$ is the topic sampled from $\theta$ for word $W$. $\beta$ is a Markov matrix giving the word distribution per topic, and $\eta$ is the Dirichlet prior parameter used in generating that matrix.

## 3.2 Dynamic Topic Models

As our preliminary experiments with historical data on both epizootic disease outbreaks and meth lab seizures showed, news flashes do not admit the kind of stationarity assumed in Figure 3. Specifically, the latent variables of our topic model change over time as a result of concept drift and the arrival of new topics, which we can think of as a birth-death process tied to observable events. Blei and Lafferty (2006) proposed a dynamic topic model with fixed topic count $K$ in which each topic's word distribution and popularity are linked over time. [5] Meanwhile, older topic modeling algorithms such as Latent Semantic Analysis (LSA) [6] that permit $K$ to vary suffer from problems such as proximity of different senses of a polysemous word, while variants Probabilistic Latent Semantic Analysis (PLSA) [7] exhibit parameter growth linear in the number of documents $D$.

### 3.2.1 Discrete Time, Infinite Topic

Ahmed and Xing (2010) proposed a partial solution to this problem by introducing an infinite Dynamic Topic Model (iDTM) that allows for an unbounded number of topics and an evolving representation of topics according to a Markovian dynamics. [8] They analyzed the birth and evolution of topics in the neural computation community based on the *Neural Information Processing Systems (NIPS)* conference proceedings. Their model evolved topics over discrete epochs (time units). All proceedings of a conference meeting fall into the same epoch. This model does not suit less "bursty topic" applications such as meth lab seizures or disease outbreaks, which are asynchronous whether reported in the news or in local law enforcement records.

For topic modeling applications such as event visualization, such discrete time model may be too brittle. An extension to continuous time will give it the needed flexibility to account for variability and change in temporal granularity.

### 3.2.2 Continuous Time, Finite Topic

Meanwhile, Wang *et al.* (2008) proposed a continuous time dynamic topic model that uses Brownian motion to simulate the evolution of topics over time. [9] Although this model uses a novel, sparse variational Kalman filtering algorithm for fast inference, the number of topics it samples from is bounded, which severely limits its application in news feed storyline creation and article aggregation. When the number of topics covered by the news feed is fewer than the pre-tuned number of topics $K$ specified in the model, similar stories will appear under different headlines. On the other hand, if the number of topics covered becomes greater than the preset number of topics, topics and headlines will get conflated.

### 3.2.3 Proposed: Continuous Time, Infinite Topic

To accommodate the needs of non-bursty news updates in domains such as our event visualization domains, we propose a hybridization of the infinite topic model and the continuous time model that combines a hierarchical Dirichlet process (DP) for dynamic topic abstraction and refinement with Brownian motion to capture stochastic topic drift. [10]

We can use a variational inference algorithm such as variational Kalman filtering to factorize the variational distribution over latent variables:

$$q(\beta_{1:T, z_{1:T}, 1:N}, \theta_{1:T} | \hat{\beta}, \phi, \gamma) = \\ \prod_{k=1}^{K} q(\beta_{1,k}, \ldots, \beta_{T,k} | \hat{\beta}_{1,k}, \ldots, \hat{\beta}_{T,k}) \times \\ \prod_{t=1}^{T} \left( q(\theta_t | \gamma_t) \prod_{n=1}^{N_t} q(z_{t,n} | \phi_{t,n}) \right)$$

where $\beta$ is the word distribution over topics and $\beta_{1:T, z_{1:T}, 1:N}$ is the word distribution over topics for time 1:$T$, topic $z_{1:T}$ and word index 1:$N$, where $N$ is the size of the document lexicon.

## 4. APPLICATION TEST BED
## 4.1 Prior Work: Veterinary Epidemiology

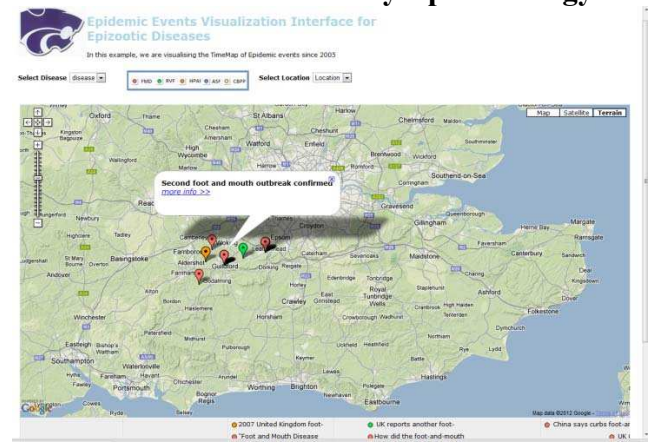

**Figure 4. Kansas Information Integration and Analysis Center (KIIAC) for epizootic diseases.**

Volkova & Hsu (2010) describes earlier work on computational information and knowledge management (CIKM) and information extraction. [11] This research, motivated by a need to visualize digests of news articles on animal disease outbreaks as shown in Figure 4, led to the earliest prototype of our event visualization system, implemented using *Google Maps* and *MIT SIMILE*. This system used syntactic detectors for semantic equivalence assertions (Volkova *et al.*, 2010). [12]

### 4.2 Kansas Meth Lab Seizures

A more recent version of the event visualization system is represented in the meth lab application described in Sections 1 and 2. This system forms the test bed for both the visualization techniques described in Section 2 and the topic modeling techniques intended to raise the precision of the federated system by improving relevance filtering and ranking.

## 5. EXPERIMENTAL EVALUATION

Figure 5 shows some baseline descriptive statistics for meth lab seizures from the test bed discussed in Sections 2 and 4.2.

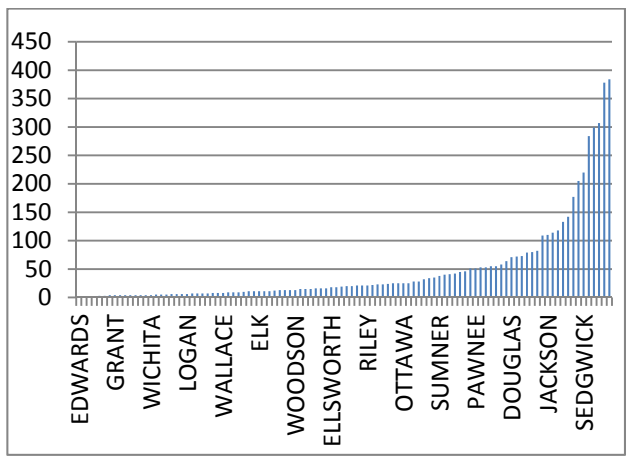

**Figure 5. Column graph of the 4942 total meth lab seizures in Kansas, 2000 - 2011, by county (104 with seizures).**

In our topic model, topics are event types, which are 50 different types of methamphetamine lab seizures. Given the text of a lab seizure report or a news story of the event with its date and location as a prior, we can use the topic model shown in Figure 3 to evaluate the likelihood of the event given the prior. Events with likelihood value above a threshold are considered highly likely to occur given the location and time of the event.

**Table 1. Topic "Abandoned dump site" proportion per seizure reports for four Kansas counties.**

|          | 2000   | 2002   | 2004   | 2006   | 2008   | 2010   |
|----------|--------|--------|--------|--------|--------|--------|
| Cowley   | **0.0345** | 0.0188 | 0.0188 | 0.0182 | 0.0001 | 0.0175 |
| Crawford | 0.0185 | 0.0172 | 0.0188 | 0.0185 | 0.0182 | 0.0188 |
| Cherokee | 0.0185 | 0.0175 | 0.0175 | 0.0178 | 0.0181 | **0.0333** |
| Reno     | **0.0350** | 0.0172 | **0.0344** | 0.0166 | **0.0527** | 0.0172 |

Given a collection of police lab seizure reports with date and location, we ran the topic model on it and evaluated the topic composition of each report in the collection. Table 1 shows the topic "Abandoned dump site" proportion per seizure reports for four Kansas counties over six years. If we set a likelihood threshold value of 0.02, then for year 2000 the event will be marked on the map for Cowley and Reno counties. The event will not get marked on the map for year 2002 for any of these four counties, and for year 2004 will be marked only in Reno County, and so on.

## 6. CONTINUING AND FUTURE WORK

In continuing work, we are validating the baseline LDA output by using it to filter and rank search results the seizure database represented in Figure 5 and Table 1. Relevance feedback from multiple subject matter experts is being used to evaluate both LDA and DP-based topic models. [11]

## 7. ACKNOWLEDGMENTS

Thanks to Loretta Wyrick-Severin (Kansas Bureau of Investigation) for assistance with public information requests and to Surya Teja Kallumadi, Tim Weninger, Svitlana Volkova, John Drouhard, Landon Fowles, and Andrew Berggren for development work on *TimeMap*.